\documentstyle[stwol]{article}

\newcommand{\np}[3]{{\sl Nucl. Phys.} {\bf #1} (19#2)~#3}
\newcommand{\pl}[3]{{\sl Phys. Lett.} {\bf #1} (19#2) #3}
\newcommand{\pr}[3]{{\sl Phys. Rev.} {\bf #1} (19#2) #3}

\newcommand{\ijm}[3]{{\sl Int. J. Mod. Phys.} {\bf #1} (19#2) #3}

\newcommand{\gsim}{\raisebox{-0.13cm}{~\shortstack{$>$ \\[-0.07cm] $\sim$}}~}

\begin{document}

\title{THEORETICAL ASPECTS OF GAUGE-BOSON PRODUCTION \\ AT LEP2 AND THE NLC}

\author{ W. BEENAKKER }

\address{Instituut--Lorentz, P.O. Box 9506, NL--2300 RA Leiden, 
         The Netherlands}


\twocolumn[\maketitle\abstracts{A report is given on the theoretical aspects
of gauge-boson production processes at LEP2 and the next linear collider
(NLC). A short discussion is given on the gauge-boson-related physics issues 
that play a role at both coliders and on the outcome of the various LEP2/NLC 
studies that have taken place during the last few years. The main emphasis of 
this report is on the question of preserving gauge-invariance when dealing 
with unstable gauge bosons. In this context a strategy is proposed for 
treating radiative corrections.}]

\section{Physics issues at LEP2}

Recently LEP2, the second stage of the LEP program, has started operation. 
After a first short run around 130\,GeV, the energy has been increased to the 
nominal threshold for the production of W-boson pairs. This opens the 
possibility of accurately measuring the W-boson mass $M_{_W}$. Two methods are 
advocated for this measurement. 
The first procedure involves a measurement of the total 
W-pair cross-section $\sigma (e^+e^- \to W^+W^- \to 4f+n\gamma)$ at an energy
close to 161\,GeV, i.e., just above the nominal threshold. In this energy 
region the cross-section $\sigma(e^+e^- \to 4f+n\gamma)$ is very sensitive 
to $M_{_W}$ through phase-space. Since for this measurement a precise knowledge
of the total cross-section is required, this method is model dependent. During 
last year's LEP2 working-group studies the conclusion was reached that an 
experimental error on the cross-section of the order of 5--6\% is feasible (for
an integrated luminosity of 50\,$\mathrm{pb}^{-1}/\mathrm{exp}$).~\cite{LEP2MW}
This translates into an envisaged experimental error of about 100\,MeV on the 
W-boson mass. From the theoretical side a systematic error of the order of
2\% is expected.~\cite{LEP2MW,LEP2WW} This uncertainty originates from 
the implementation of initial-state (photon) radiation (ISR), from
missing ${\cal O}(\alpha)$ corrections, and from the dependence of the 
cross-section on the unknown Higgs-boson mass~\cite{MHatWWthr}. 
The second procedure is a model-independent one. It consists in a 
reconstruction of the Breit--Wigner resonance shapes of the W bosons from the 
measured momenta of the decay products. For this method higher energies 
($\gsim 175$\,GeV) are required in order to have a sufficiently large 
W-pair production rate. The expected experimental error on the 
reconstructed W-boson mass is 40--50\,MeV for an integrated luminosity of 
500\,$\mathrm{pb}^{-1}/\mathrm{exp}$.~\cite{LEP2MW} From the theoretical side 
two main sources of systematic uncertainties can be identified. First of all
a precise knowledge of ISR is required, in view of the fact that the W-pair 
energy will be different from the laboratory energy of the incoming electrons 
and positrons in the presence of energy losses in the initial state. 
This particular uncertainty seems to be under control, amounting to 
roughly 10\,MeV from estimates of the average energy loss.~\cite{LEP2WW} In the
hadronic--leptonic decay channels this is the main uncertainty. In the purely 
hadronic (four-jet) decay channels, fragmentation effects can add another 
50\,MeV to the uncertainty.~\cite{LEP2MW}  As the average space--time distance 
between the two decaying W bosons is smaller than 0.1\,fm, i.e., less than the 
typical hadronic size of 1\,fm, the fragmentation of the two W bosons is not 
independent (``color rearrangement''). In view of the overlapping hadronization
regions also coherence effects are possible between identical low-momentum 
bosons stemming from different W bosons (``Bose--Einstein correlations''). 
Both fragmentation effects are strongly model-dependent, so further theoretical
and experimental studies are needed before conclusive statements can be made.\\
\indent
The envisaged precision of 40--50\,MeV in the determination of $M_{_W}$ at 
LEP2 constitutes a significant improvement on the present hadron-collider
measurements ($\Delta M_{_W} \approx 125$\,MeV)~\cite{MWdirectWarsaw}. Since 
the mass of the W boson is one of the key parameters of the
electroweak theory, such an improved accuracy makes the tests on the Standard 
Model (SM) of electroweak interactions more stringent. In order to facilitate 
the extraction of this piece of experimental information, a new 
input-parameter scheme was introduced.~\cite{LEP2WW} 
Besides the usual LEP1 input parameters: $\alpha$, $G_\mu$, $M_{_Z}$ and the 
light-fermion masses, also $M_{_W}$ is treated as input (fit parameter).
From muon decay the mass of the top quark can be calculated as a function of
the Higgs-boson mass $M_{_H}$ and the strong coupling $\alpha_s$:
\begin{displaymath}
  G_\mu = \frac{\alpha\pi/\sqrt{2}}{M_{_W}^2\,(1-M_{_W}^2/M_{_Z}^2)}\,
          \frac{1}{1-\Delta r(m_t,M_{_H},\alpha_s)}.
\end{displaymath}
The so-obtained top-quark mass can then be confronted with the direct bounds
from the Tevatron and the indirect ones from the precision measurements at 
LEP1/SLC. In this way improved limits on $M_{_H}$ can be obtained.

The second piece of information that LEP2 can provide is the structure of the 
triple gauge-boson couplings (TGC). These couplings appear at tree level in 
LEP2 processes like $e^+e^- \to 4f$ or $e^+e^- \to 2f+\gamma$,
in contrast to LEP1 where they only entered through loop corrections. The 
largest experimental sensitivity to the TGC is achieved by going to the 
highest available energy and by investigating angular distributions of the 
gauge bosons and their subsequent decay products. The theoretical uncertainties
associated with these distributions are estimated to be 1--2\%, originating 
from the implementation of ISR and from missing ${\cal O}(\alpha)$ 
corrections.~\cite{LEP2WW} Assuming an energy of 
190\,GeV and an integrated luminosity of 500\,$\mathrm{pb}^{-1}/\mathrm{exp}$, 
a determination at the level of $\Delta(TGC)= 0.05$--0.1 seems 
feasible~\cite{LEP2TGC}, provided one applies symmetry or operator-hierarchy 
arguments to reduce the number of independent couplings (to one or two at 
most). Based on these studies it is safe to say that the Yang--Mills 
character of the TGC can be established at LEP2. Since the TGC are at the 
heart of non-abelian theories, this information is essential for a direct 
confirmation of the SM or for providing a window to physics beyond the SM.

\section{Physics issues at the NLC}

The afore-mentioned LEP2 TGC studies can much more efficiently be performed
at the NLC. This is based on the notion that couplings different from the 
SM Yang--Mills ones, called anomalous or non-standard couplings, will in
general upset the intricate gauge cancelations for longitudinal gauge bosons 
and lead to a high-energy behavior of the cross-sections increasing with 
energy and thereby violating unitarity. At LEP2 the energy is too low to see 
such effects, but at the NLC, with its energy in the range 500--2000\,GeV,
this feature can be fully exploited. Combined with the increased luminosity at 
the NLC this allows a determination of the TGC below the percent 
level~\cite{NLCTGC}, i.e., in the range predicted by specific models. Moreover,
the $e\gamma$ and $\gamma\gamma$ modes of the NLC lead to an increased 
sensitivity to individual couplings (in processes like $e\gamma \to 3f$).

At the NLC one can go one step further. The high energies open the possibility 
of studying quartic gauge-boson couplings (QGC) in reactions like 
$e^+e^- \to 6f$ or $\gamma\gamma \to 4f$, thereby entering the realm of the 
symmetry-breaking mechanism. On top of that one can look for signs of a 
strongly-interacting symmetry-breaking sector (i.e., resonances or phase 
shifts), by studying longitudinal gauge-boson interactions (rescattering) in 
$e^+e^- \to 4f,\,6f$ and $\gamma\gamma \to 4f$. In other words, one of the 
most important tasks to be performed at the NLC is a detailed investigation of 
the Higgs sector or any alternative thereof.

\section{Gauge-invariant treatment of unstable gauge bosons}

\subsection{Lowest order}

The above-described physics issues all involve an investigation of processes
with photons and/or fermions in the initial and final state.
If complete sets of graphs contributing to a given process are taken into
account, the associated matrix elements are in principle gauge-invariant.
However, the gauge bosons that appear as intermediate particles can give
rise to poles $1/(k^2-M^2)$ if they are treated as stable particles. This can 
be cured by introducing the finite decay width in one way or another, while at 
the same time preserving gauge independence and, through a proper high-energy 
behavior, unitarity. In field theory, such widths arise naturally from the 
imaginary parts of higher-order diagrams describing the gauge-boson 
self-energies, resummed to all orders. This procedure has been used with great 
success in the past: indeed, the $Z$ resonance can be described to very high 
numerical accuracy. However, in doing a Dyson summation of self-energy graphs, 
we are singling out only a very limited subset of all the possible higher-order
diagrams. It is therefore not surprising that one often ends up with a result 
that retains some gauge dependence. 

Till recently two approaches were popular
in the construction of lowest-order LEP2/NLC Monte Carlo generators. The first 
one involves the systematic replacement $1/(k^2-M^2) \to 1/(k^2-M^2+iM\Gamma)$,
also for $k^2<0$. Here $\Gamma$ denotes the physical width of the gauge boson
with mass M and momentum $k$. This scheme is called the `fixed-width scheme'.
As in general the resonant diagrams are not gauge-invariant by themselves, 
this substitution will destroy gauge invariance. Moreover, it has no 
physical motivation, since in perturbation theory the propagator for 
space-like momenta does not develop an imaginary part. Consequently, 
unitarity is violated in this scheme. To improve on the latter another 
approach can be adopted, involving the use of a running width $iM\Gamma(k^2)$ 
instead of the constant one $iM\Gamma$ (`running-width scheme'). This,
however, still cannot cure the problem with gauge invariance. 

At this point one might ask oneself the 
legitimate question whether the gauge-breaking terms are numerically relevant 
or not. After all, the gauge breaking is caused by the finite decay width and
is, as such, in principle suppressed by powers of $\Gamma/M$. From LEP1 we 
know that gauge breaking can be negligible for all practical purposes. 
However, the presence of small scales can amplify the gauge-breaking terms.
This is for instance the case for almost collinear space-like photons or
longitudinal gauge bosons at high energies, involving scales of 
${\cal O}(p_{_B}^2/E_{_B}^2)$ (with $p_{_B}$ the 
momentum of the involved gauge boson). In these situations the external 
current coupled to the photon or to the longitudinal gauge boson becomes 
approximately proportional to $p_{_B}$. In other words, in these 
regimes sensible theoretical predictions are only possible if the amplitudes 
with external currents replaced by the corresponding gauge-boson momenta 
fulfill appropriate Ward identities.

In order to substantiate these statements, a truly gauge-invariant scheme is
needed. It should be stressed, however, that any such scheme is arbitrary to a 
greater or lesser extent: since the Dyson summation must necessarily be taken 
to all orders of perturbation theory, and we are not able to compute the 
complete set of {\it all} Feynman diagrams to {\it all} orders, the various 
schemes differ even if they lead to formally gauge-invariant results. Bearing 
this in mind, we need some physical motivation for choosing a particular 
scheme. In this context two options can be mentioned, which fulfill the 
criteria of gauge invariance and physical motivation. 
The first option is the so-called `pole scheme'.~\cite{Veltman,Stuart,Aeppli} 
In this scheme one decomposes the complete amplitude according to the pole 
structure by expanding around the poles 
(e.g.~$f(k^2)/(k^2-M^2) = f(M^2)/(k^2-M^2) + \mathrm{finite~terms}$). As the 
physically observable residues of the poles are gauge-invariant, gauge
invariance is not broken if the finite width is taken into account in the pole 
terms $\propto 1/(k^2-M^2)$. It should be noted, however, that there exists 
some controversy in the literature about the `correct' procedure for doing 
this and about the range of validity of the pole scheme, especially in the 
vicinity of thresholds. 
The second option is based on
the philosophy of trying to determine and include the minimal set of Feynman 
diagrams that is necessary for compensating the gauge violation caused by the 
self-energy graphs. This is obviously the theoretically most satisfying 
solution, but it may cause an increase in the complexity of the matrix 
elements and a consequent slowing down of the numerical calculations. For the 
gauge bosons we are guided by the observation that the lowest-order decay 
widths are exclusively given by the imaginary parts of the fermion loops in 
the one-loop self-energies. It is therefore natural to perform a Dyson 
summation of these fermionic one-loop self-energies and to include the other 
possible one-particle-irreducible fermionic one-loop corrections 
(``fermion-loop scheme'').~\cite{BHF1} For the LEP2 process $e^+e^- \to 4f$ 
this amounts to adding the fermionic triple gauge-boson vertex corrections. 
The complete set of fermionic contributions forms a gauge-independent subset 
and obeys all Ward identities exactly, even with resummed 
propagators.~\cite{BHF2} 
As mentioned above, the validity of the Ward identities guarantees a proper 
behavior of the cross-sections in the presence of collinear photons and at 
high energies in the presence of longitudinal gauge-boson modes. On top of 
that, within the fermion-loop scheme the appropriately renormalized matrix 
elements for the generic LEP2 process $e^+e^- \to 4f$ can be formulated in 
terms of effective Born matrix elements, using the familiar language of 
running couplings.~\cite{BHF2}

A numerical comparison of the various schemes~\cite{BHF1,BHF2} confirms the 
importance of not violating the Ward identities. For the LEP2 process 
$e^+e^- \to e^-\bar{\nu}_e\,u\bar{d}$, a process that is particularly 
important for TGC studies, the impact of violating the $U(1)$ electromagnetic
gauge invariance was demonstrated.~\cite{BHF1} Of the above-mentioned schemes 
only the running-width scheme violates $U(1)$ gauge invariance. The associated 
gauge-breaking terms are enhanced in a disastrous way by a factor of 
${\cal O}(s/m_e^2)$, in view of the fact that the electron may emit a virtual
(space-like) photon with $p_\gamma^2$ as small as $m_e^2$. A similar 
observation can be made at high energies (NLC) when some of the 
intermediate gauge bosons become effectively longitudinal. There too the 
running-width scheme renders completely unreliable results.~\cite{BHF2} In 
processes involving more intermediate gauge bosons, e.g.~$e^+e^- \to 6f$,
also the fixed-width scheme breaks down at high energies as a result of
breaking $SU(2)$ gauge invariance.

\subsection{Radiative corrections}

By employing the fermion-loop scheme all one-particle-irreducible fermionic 
one-loop corrections can be embedded in the tree-level matrix elements. This 
results in running couplings, propagator functions, vertex functions, etc. 
However, there is still the question about the bosonic corrections. A large 
part of these bosonic corrections, as e.g.~the leading QED corrections, 
factorize and can be treated by means of a convolution, using the 
fermion-loop-improved cross-sections in the integration kernels. This allows 
the inclusion of higher-order QED corrections and soft-photon 
exponentiation. In this way various important effects can be covered, 
as e.g.~the large negative soft-photon corrections near the nominal W-pair 
threshold, the distortion of angular distributions as a result of hard-photon 
boost effects, and the average energy loss due to radiated 
photons.~\cite{LEP2WW,WWreview} Nevertheless, the remaining bosonic corrections
can be large, especially at high energies.~\cite{LEP2WW,WWreview}

In order to include these corrections one might attempt to extend the 
fermion-loop scheme. In the context of the background-field method a Dyson 
summation of bosonic self-energies can be performed without violating
the Ward identities.~\cite{BFM} However, the resulting matrix elements depend
on the quantum gauge parameter at the loop level that is not completely 
taken into account. As mentioned before, the perturbation series has to be 
truncated; in that sense the dependence on the quantum gauge parameter could be
viewed as a parametrization of the associated ambiguity. 

As a more appealing strategy one might adopt a hybrid scheme, adding the 
remaining bosonic loop corrections by means of the pole scheme. This is 
gauge-invariant and contains the well-known bosonic corrections for the 
production of on-shell gauge bosons (in particular W-boson pairs). Moreover, 
if the quality of the pole scheme were to degrade in certain regions of 
phase-space, the associated error is reduced by factors of $\alpha/\pi$. 
It should be noted that the application of the pole scheme to photonic 
corrections requires some special care, because in that case terms 
proportional to $\log(k^2-M^2)/(k^2-M^2)$ complicate the pole
expansion.~\cite{Aeppli,WWreview}

\section*{Acknowledgment}
This research has been supported by a fellowship of the Royal Dutch Academy of
Arts and Sciences.

\end{document}